\documentclass[a4paper,12pt]{article}
\usepackage{graphicx, color}
\usepackage{amsmath, amsfonts, amssymb, bm}
\newcommand{\p}{\, .}
\newcommand{\sv}{\, ,}
\newcommand{\eq}[1]{(\ref{#1})}
\newcommand{\dm}{\displaystyle}
\author{Armando Majorana
\\[5pt] Dipartimento di Matematica e Informatica
\\ Viale A. Doria 6, 95125 Catania, Italy}
\title{}
\title{Stationary solutions to a Vlasov equation for planetary rings.}
\begin{document}
\maketitle
%
%
\begin{abstract}
In this paper we consider a Vlasov or collisionless Boltzmann equation describing the dynamics of 
planetary rings. 
We propose a simple physical model, where the particles of the rings move under the gravitational
Newtonian potential of two primary bodies.
We neglect the gravitational forces between the particles.
We use a perturbative technique, which allows to find explicit solutions at the first 
order and solutions at the second order, solving a set of two linear ordinary differential 
equations.
\end{abstract}
{\bf Keywords}
Planetary rings, Vlasov equation, Stationary explicit solutions,
Circular restricted three-body problem.
\clearpage
\section{Introduction and basic equations}
The gravitational $N$-body problem is one of the oldest problems in physics. 
The $N$ bodies interact classically through Newton’s Law of Universal Gravitation. 
Then the equations of motion are
\begin{equation}
m_{i} \, \ddot{\mathbf{r}}_{i} = \mbox{} - G \sum_{j=1, j \neq i}^{N} m_{j} \, 
\dfrac{\mathbf{r}_{i} - \mathbf{r}_{j}}{\left| \mathbf{r}_{i} - \mathbf{r}_{j} \right|^{3}}
\sv \quad (i =1, 2, ..., N)
\label{eqN}
\end{equation}
where $m_{i}$ is the mass of the body $P_{i}$, $\mathbf{r}_{i}$ is its position vector relative 
to some inertial frame, and $G$ is the universal constant of gravitation. 
These equations provide a reasonable and well-accepted mathematical model with numerous 
applications in astrophysics, including the motion of planets, asteroids, comets and other bodies in 
the Solar System. \\
The number $N$ of the bodies can be very large; for instance, the planetary rings are composed of a 
large number of small bodies with sizes from specks of dust to small moons. Saturn’s rings are the 
largest and best studied. \\
In general, for huge $N$, an alternative approach consists to consider a statistical description 
through a kinetic (Boltzmann) equation \cite{Binney}. 
Unfortunately, Boltzmann equation is a non-linear integro-differential equation and solving it in 
six-dimensional phase space requires an extremely large memory and computational time. 
So, alternative kinetic equations, as collisionless Boltzmann equation
\cite{Rein}, Bhatnagar-Gross-Krook (BGK) models, where a relaxation term replaces the collision
integral, and other models \cite{Severne} are been considered.
Very recently, accurate numerical solutions to the Vlasov-Poisson model for self-gravitating systems
are proposed in \cite{Cheng} and \cite{Yoshikawa}.
The literature is very rich in papers devoted to analytical, numerical and computational studies on
this topic. 

In this paper we will restrict our attention to a Vlasov or collisionless Boltzmann equation
describing the dynamics of planetary rings. 
The stability and the structure of Saturn's rings was studied by Griv et al.
\cite{Griv2000}-\cite{Griv2003c} using both collisionless and BGK models.

A simple mathematical model describing the dynamics of a planetary rings is given by the following 
$N$ circular restricted three body problems.\\
\emph{
A large set of small bodies, are subject to the attraction of the Sun and a planet. The primary 
bodies move in a plane in circular orbits about their center of mass. The total mass of the small 
bodies is negligible compared to the primary body masses. Then the presence of the small bodies 
does not disturb the circular motion of the two large bodies.
}
\\
We denote by $\mathbf{r}_{S}$ and $m_{S}$ the position and the mass of the Sun; 
$\mathbf{r}_{P}$ and $m_{P}$, are the position a the mass of the planet.
Hence, the equations of motion are
\begin{equation}
m_{i} \, \ddot{\mathbf{r}}_{i} = \mbox{} - G \left[
 \dfrac{m_{i} \, m_{S}}{| \mathbf{r}_{i} - \mathbf{r}_{S} |^{3}} 
 \left( \mathbf{r}_{i} - \mathbf{r}_{S} \right) +
 \dfrac{m_{i} \, m_{P}}{| \mathbf{r}_{i} - \mathbf{r}_{P} |^{3}} \, 
 \left( \mathbf{r}_{i} - \mathbf{r}_{P} \right)  \right] 
 \quad (i =1, 2, ..., N) \p
\label{eqSPp} 
\end{equation}
Since we are considering a large number of small bodies moving under the gravitational potential 
of the primary bodies, at any time $t$ a full description of the state of this system can be
given by specifying the number of small bodies 
$f(t, \mathbf{r}, \bm{\xi}) \, d \mathbf{r} \, d \bm{\xi}$,
having positions in the small volume $d \mathbf{r}$ centered on $\mathbf{r}$ in the small velocity 
range $d \bm{\xi}$ centered on $\bm{\xi}$. 
The function  $f(t, \mathbf{r}, \bm{\xi})$ is called the \emph{distribution function} of the 
system. 
Obviously, we require that $f \geq 0$ almost everywhere, since we do not allow particles with 
negative mass.\\
We assume that the following Vlasov equation describes the evolution of the distribution 
function $f(t, \mathbf{r}, \bm{\xi})$
\begin{equation}
\dfrac{\partial f}{\partial t} + \bm{\xi} \cdot \nabla_{\mathbf{r}} f -
\nabla_{\mathbf{r}} \Phi \cdot \nabla_{\bm{\xi}} f = 0 \sv
\label{eqVxi}
\end{equation}
where
\begin{equation}
\Phi(t, \mathbf{r}) = \mbox{} - G \left[ \dfrac{m_{S}}{| \mathbf{r} - \mathbf{r}_{S} |} +
\dfrac{m_{P}}{| \mathbf{r} - \mathbf{r}_{P} |} \right] 
\label{potSP}
\end{equation}
is gravitational potential of the primary bodies. \\
We denote by $(X,Y,Z)$ the component of the vector $\mathbf{r}$. 
Now, we introduce a uniformly rotating coordinate system with origin at the mass center of the 
primary bodies, so that the Sun and the planet are located on the $x$ axis with coordinates 
$(x_{S},0,0)$ with 
$x_{S} > 0$, and $(x_{P},0,0)$, respectively.
This implies the following transformation of variables 
\begin{equation}
\left\{
\begin{array}{rcl}
 X & = & x \, \cos(\omega t) - y \, \sin(\omega t)
\\[0pt]
 Y & = & x \, \sin(\omega t) + y \, \cos(\omega t)
\\[0pt]
 Z & = & z
\\[0pt]
 \xi_{1} & = & c_{1} \, \cos(\omega t) - c_{2} \, \sin(\omega t)
 - \omega \, x \, \sin(\omega t) - \omega \, y \, \cos(\omega t) 
\\[0pt]
 \xi_{2} & = & c_{1} \, \sin(\omega t) + c_{2} \, \cos(\omega t)
 + \omega \, x \, \cos(\omega t) - \omega \, y \, \sin(\omega t) 
\\[0pt]
 \xi_{3} & = & c_{3}
\end{array}
\right. \label{trasf}
\end{equation}
where $(x,y,z)$ are the new spatial coordinates, $(c_{1},c_{2},c_{3})$ the component of the 
particle's velocity and $\omega$ is the constant angular velocity of the primary bodies.
It is useful to introduce cylindrical coordinates (centered in $P$) given by
\begin{equation}
\left\{
\begin{array}{rcl}
 x & = & x_{P} + r \, \cos \theta
\\[0pt]
 y & = & r \, \sin \theta
\\[0pt]
 z & = & z
\\[0pt]
 c_{1} & = & u_{r} \, \cos \theta - u_{\theta} \, \sin \theta 
\\[0pt]
 c_{2} & = & u_{r} \, \sin \theta + u_{\theta} \, \cos \theta 
\\[0pt]
 c_{3} & = & u_{z}
\end{array}
\right. \label{trasf_cil}
\end{equation}
In terms of the new variables, the distribution function $f(t, \mathbf{r}, \bm{\xi})$ is
replaced by the new unknown ${\cal{G}}(t, r, \theta, z, u_{r}, u_{\theta}, u_{z}) $, 
and Eq.\eq{eqVxi} writes
\begin{eqnarray}
&& \dm 
\dfrac{\partial \cal{G}}{\partial t} + 
u_{r} \, \dfrac{\partial \cal{G}}{\partial r} +
\dfrac{u_{\theta}}{r} \, \dfrac{\partial \cal{G}}{\partial \theta} +
u_{z} \, \dfrac{\partial \cal{G}}{\partial z} +
\left[
\omega^{2} \left( x_{P} \, \cos \theta + r \right) + 2 \, \omega \, u_{\theta} +
\dfrac{u_{\theta}^{2}}{r} - \dfrac{\partial V}{\partial r}
\right] \dfrac{\partial \cal{G}}{\partial u_{r}} \nonumber
\\[15pt]
&& \dm \mbox{} \quad -
\left[
\omega^{2} \, x_{P} \, \sin \theta + 2 \, \omega \, u_{r} +
\dfrac{u_{r} \, u_{\theta}}{r} + \dfrac{1}{r} \, \dfrac{\partial V}{\partial \theta}
\right] \dfrac{\partial \cal{G}}{\partial u_{\theta}} -
\dfrac{\partial V}{\partial z} \, \dfrac{\partial \cal{G}}{\partial u_{z}} = 0 \sv
\label{eqVrt}
\end{eqnarray}
where, now, the gravitational potential is
\begin{equation}
V(r, \theta, z) = \mbox{} - G \left[ \dfrac{m_{S}}
{ \sqrt{ (x_{P} - x_{S})^{2} + r^{2} - 2 \, |x_{P} - x_{S}| \, r \, \cos \theta + z^{2} } }
+ \dfrac{m_{P}}{ \sqrt{ r^{2} + z^{2} } } \right] .
\label{VSP}
\end{equation}
We will study the kinetic model given by Eq.~\eq{eqVrt} and Eq.~\eq{VSP}.
\section{The 2D equations and an approximate model}
Assuming that all the particles move on the plan of the primary bodies, Eqs.~\eq{eqVrt}-\eq{VSP}
reduce to
\begin{eqnarray}
&& 
\dfrac{\partial \cal{G}}{\partial t} + 
u_{r} \, \dfrac{\partial \cal{G}}{\partial r} +
\dfrac{u_{\theta}}{r} \, \dfrac{\partial \cal{G}}{\partial \theta} +
\left[
\omega^{2} \left( x_{P} \, \cos \theta + r \right) + 2 \, \omega \, u_{\theta} +
\dfrac{u_{\theta}^{2}}{r} - \dfrac{\partial V}{\partial r}
\right] \dfrac{\partial \cal{G}}{\partial u_{r}} \nonumber
\\[15pt]
&& \mbox{} \quad -
\left[
\omega^{2} \, x_{P} \, \sin \theta + 2 \, \omega \, u_{r} +
\dfrac{u_{r} \, u_{\theta}}{r} + \dfrac{1}{r} \, \dfrac{\partial V}{\partial \theta}
\right] \dfrac{\partial \cal{G}}{\partial u_{\theta}} = 0 \sv
\label{eqV2rt}
\end{eqnarray}
with
\begin{equation}
V(r, \theta, z) = \mbox{} - G \left[ \dfrac{m_{S}}
{ \sqrt{ (x_{P} - x_{S})^{2} + r^{2} - 2 \, |x_{P} - x_{S}| \, r \, \cos \theta } }
+ \dfrac{m_{P}}{r} \right] .
\label{V2SP} 
\end{equation}
If we consider the Saturn's rings, the mean distance of rings from the center of the planet is
small with respect to the distance Sun-Saturn. 
Hence, it is reasonable to replace the exact potential due to the Sun, with a polynomial
approximation, obtained using Mac Laurin expansion.
We have
$$
\dfrac{1}
{ \sqrt{ (x_{P} - x_{S})^{2} + r^{2} - 2 \, |x_{P} - x_{S}| \, r \, \cos \theta } } 
\approx
\dfrac{1}{|x_{P} - x_{S}|} +
\dfrac{\cos \theta}{|x_{P} - x_{S}|^{2}} \, r +
\dfrac{3 \, \cos^{2} \theta - 1}{2 \, |x_{P} - x_{S}|^{3}} \, r^{2} \sv
$$
and
\begin{equation}
V(r, \theta, z) \approx
\mbox{} - G \, m_{S} \left[ \dfrac{1}{|x_{P} - x_{S}|} +
\dfrac{\cos \theta}{|x_{P} - x_{S}|^{2}} \, r +
\dfrac{3 \, \cos^{2} \theta - 1}{2 \, |x_{P} - x_{S}|^{3}} \, r^{2} \right]
- \dfrac{G \, m_{P}}{r} \p
\label{Vapprox}
\end{equation}
Using Eq.\eq{Vapprox}, Vlasov equation \eq{eqV2rt} becomes
\begin{eqnarray}
&& \mbox{} \hspace{-35pt} 
\dfrac{\partial \cal{G}}{\partial t} + 
u_{r} \, \dfrac{\partial \cal{G}}{\partial r} +
\dfrac{u_{\theta}}{r} \, \dfrac{\partial \cal{G}}{\partial \theta} +
\left[
\omega^{2} \left( x_{P} \, \cos \theta + r \right) + 2 \, \omega \, u_{\theta} +
\dfrac{u_{\theta}^{2}}{r} \right. \nonumber
\\[15pt]
&& \mbox{} \hspace{-30pt} + \left.
 G \, m_{S} \left( \dfrac{\cos \theta}{|x_{P} - x_{S}|^{2}} 
 + \dfrac{3 \, \cos^{2} \theta - 1}{|x_{P} - x_{S}|^{3}} \, r
\right) - \dfrac{G \, m_{P}}{r^{2}}
\right] \dfrac{\partial \cal{G}}{\partial u_{r}} \nonumber
\\[15pt]
&& \dm \mbox{} \hspace{-30pt} -
\left[
\omega^{2} \, x_{P} \, \sin \theta + 2 \, \omega \, u_{r} +
\dfrac{u_{r} \, u_{\theta}}{r} + 
 G \, m_{S} \left( \dfrac{\sin \theta}{|x_{P} - x_{S}|^{2}}
 + \dfrac{3 \, \cos \theta \, \sin \theta}{|x_{P} - x_{S}|^{3}} \, r \right)
\right] \dfrac{\partial \cal{G}}{\partial u_{\theta}} = 0 \p
\label{eqV2app}
\end{eqnarray}
Since
$$
\omega^{2} \, x_{P} = \mbox{} - \dfrac{G \, m_{S}}{|x_{P} - x_{S}|^{2}} \sv
$$
if we define
$$
\varepsilon = \dfrac{G \, m_{S}}{|x_{P} - x_{S}|^{3}} 
\quad \mbox{and} \quad \mu = G \, m_{P} \sv
$$
then Eq.~\eq{eqV2app} simplifies and writes
\begin{eqnarray}
&& \mbox{}  
\dfrac{\partial \cal{G}}{\partial t} + 
u_{r} \, \dfrac{\partial \cal{G}}{\partial r} +
\dfrac{u_{\theta}}{r} \, \dfrac{\partial \cal{G}}{\partial \theta} +
\left[
\omega^{2} \, r + 2 \, \omega \, u_{\theta} +
\dfrac{u_{\theta}^{2}}{r} + \varepsilon \left( 3 \, \cos^{2} \theta - 1 \right) r
- \dfrac{\mu}{r^{2}} \right] \dfrac{\partial \cal{G}}{\partial u_{r}} 
\nonumber
\\[15pt]
&& \dm \mbox{} \quad -
\left[
2 \, \omega \, u_{r} +
\dfrac{u_{r} \, u_{\theta}}{r} + 
 \varepsilon \left( 3 \, \cos \theta \, \sin \theta \right) r 
\right] \dfrac{\partial \cal{G}}{\partial u_{\theta}} = 0 \p
\label{eqV2sim}
\end{eqnarray}
Often the parameter $\varepsilon$ is small.
For instance, in the case of Saturn's rings, using the following units of measure
\begin{center}
$5 \times 10^{5}$ km (length), 
$5.68319 \times 10^{26}$ kg (Saturn's mass) and
$3600$ s (time),
\end{center}
then the maximum distance of the rings from the center of Saturn is approximatively $0.96$, 
and we have
$$
\dfrac{\mu}{r} \ge 3.93 \times 10^{-3} \quad (\mbox{for } r < 1)
\quad \mbox{ and } \quad
\varepsilon \approx 5.92 \times 10^{-10} \p
$$
In this case, it is required a very accurate numerical scheme, which takes into account the 
effects of the small term $\varepsilon$ in Eq.~\eq{eqV2sim}, for large time integration.
In order to overcome this difficulty, we suggest a simple (Hilbert) expansion, assuming that
\begin{equation}
{\cal{G}}(t, r, \theta, u_{r}, u_{\theta}) \approx
{\cal{G}}_{0}(t, r, \theta, u_{r}, u_{\theta}) + \varepsilon \,
{\cal{G}}_{1}(t, r, \theta, u_{r}, u_{\theta})  \p
\label{expG} 
\end{equation}
Now, Eq.~\eq{eqV2sim} gives the following set of partial differential equations
\begin{eqnarray}
&&
\dfrac{\partial {\cal{G}}_{0}}{\partial t} + 
u_{r} \, \dfrac{\partial {\cal{G}}_{0}}{\partial r} +
\dfrac{u_{\theta}}{r} \, \dfrac{\partial {\cal{G}}_{0}}{\partial \theta} +
\left( \omega^{2} \, r + 2 \, u_{\theta} \, \omega +
\dfrac{u_{\theta}^{2}}{r} 
 - \dfrac{\mu}{r^{2}} \right) \dfrac{\partial {\cal{G}}_{0}}{\partial u_{r}} 
\nonumber
\\[15pt]
&& \mbox{} \quad - 
\left( 2 \, u_{r} \, \omega +
\dfrac{u_{r} \, u_{\theta}}{r}
\right) \dfrac{\partial {\cal{G}}_{0}}{\partial u_{\theta}} = 0 \sv
\label{eqG0}
\\[15pt]
&&
\dfrac{\partial {\cal{G}}_{1}}{\partial t} + 
u_{r} \, \dfrac{\partial {\cal{G}}_{1}}{\partial r} +
\dfrac{u_{\theta}}{r} \, \dfrac{\partial {\cal{G}}_{1}}{\partial \theta} +
\left( \omega^{2} \, r + 2 \, u_{\theta} \, \omega +
\dfrac{u_{\theta}^{2}}{r} 
 - \dfrac{\mu}{r^{2}} \right) \dfrac{\partial {\cal{G}}_{1}}{\partial u_{r}} 
\nonumber
\\[15pt]
&& \mbox{} \quad - 
\left( 2 \, u_{r} \, \omega + \dfrac{u_{r} \, u_{\theta}}{r} 
\right) \dfrac{\partial {\cal{G}}_{1}}{\partial u_{\theta}} 
= 
\left( 1 - 3 \, \cos^{2} \theta \right) r \, \dfrac{\partial {\cal{G}}_{0}}{\partial u_{r}}
+ \left( 3 \, \cos \theta \, \sin \theta \right) r \, 
\dfrac{\partial {\cal{G}}_{0}}{\partial u_{\theta}} \p
\label{eqG1}
\end{eqnarray}
We note the splitting of the equations, and, of course, we solve before Eq.~\eq{eqG0} and then 
Eq.~\eq{eqG1}.
It is evident that Eq.~\eq{eqG0} is the Vlasov equation for an ensemble of particles moving in the
gravitational field of a central mass.
\subsection{The equation to ${\cal{G}}_{0}$}
Here, we use the method of the characteristic curves for solving the first partial differential 
equation. 
This method allows to find analytical or numerical solutions to linear partial differential 
equations for fixed initial condition, and sometimes classes of explicit solutions.
In our case, we get a set of ordinary differential equations corresponds to Eq.~\eq{eqG0}. 
\begin{equation}
\left\{
\begin{array}{l}
\dfrac{dr}{dt} = u_{r}
\\[12pt]
\dfrac{d \theta}{dt} = \dfrac{u_{\theta}}{r}
\\[12pt]
\dfrac{d u_{r}}{dt} = \omega^{2} \, r + 2 \, u_{\theta} \, \omega + \dfrac{u_{\theta}^{2}}{r} 
 - \dfrac{\mu}{r^{2}} 
\\[12pt]
\dfrac{d u_{\theta}}{dt} = - \, 2 \, u_{r} \, \omega - \dfrac{u_{r} \, u_{\theta}}{r}
\end{array}
\right. .
\label{eqchG0}
\end{equation}
In the appendix A we give the mathematical details of the study of system \eq{eqchG0}.
Here, we show a class of exact solutions to Eq.~\eq{eqG0}.
It is easy to verify that every differentiable function
\begin{equation}
\mathbb{F}(r, u_{r}, u_{\theta}) =
\mathbb{G} \left(  r \, u_{\theta} + \omega \, r^{2} ,
\dfrac{1}{2} \left( u_{r} \right)^{2} + \dfrac{1}{2} 
\left( u_{\theta} + \omega \, r \right)^{2} - \dfrac{\mu}{r} \right)
\label{solF} 
\end{equation}
satisfies Eq.~\eq{eqG0}.
This result recall Jean's Theorem and the existence of stationary spherically symmetric 
solutions to a Vlasov equations for stellar dynamics \cite{Batt}. \\
It is useful to define
\begin{equation}
\varphi_{1}(r, u_{\theta}) = r \, u_{\theta} + \omega \, r^{2} \sv
\quad 
\varphi_{2}(r, u_{r}, u_{\theta}) = \dfrac{1}{2} \left( u_{r} \right)^{2} + \dfrac{1}{2} 
\left( u_{\theta} + \omega \, r \right)^{2} - \dfrac{\mu}{r} \p
\label{f1f2}
\end{equation}
The stationary solutions \eq{solF} do not depend on $\theta$.
Since, in Eq.~\eq{eqG0}, the gravitational potential depends only on the distance $r$ from the 
origin,  spherically symmetric solutions have a clear physical meaning.
\subsection{The equation to ${\cal{G}}_{1}$}
Using two elementary trigonometric formulas,
Eq.~\eq{eqG1} becomes
\begin{eqnarray}
&&
\dfrac{\partial {\cal{G}}_{1}}{\partial t} + 
u_{r} \, \dfrac{\partial {\cal{G}}_{1}}{\partial r} +
\dfrac{u_{\theta}}{r} \, \dfrac{\partial {\cal{G}}_{1}}{\partial \theta} +
\left( \omega^{2} \, r + 2 \, u_{\theta} \, \omega +
\dfrac{u_{\theta}^{2}}{r} 
 - \dfrac{\mu}{r^{2}} \right) \dfrac{\partial {\cal{G}}_{1}}{\partial u_{r}}
\nonumber
\\[15pt]
&& \mbox{} \quad
- \left( 2 \, u_{r} \, \omega +
\dfrac{u_{r} \, u_{\theta}}{r} 
\right) \dfrac{\partial {\cal{G}}_{1}}{\partial u_{\theta}} 
= 
- \, \dfrac{1}{2} \left( 1 + 3 \, \cos 2\theta \right) r \, 
\dfrac{\partial {\cal{G}}_{0}}{\partial u_{r}}
+ \dfrac{3}{2} \left( \sin 2\theta \right) r \, 
\dfrac{\partial {\cal{G}}_{0}}{\partial u_{\theta}} \p
\label{eqGsc}
\end{eqnarray}
If ${\cal{G}}_{0}$ does not depend on $\theta$, then we can look for solutions of the kind
\begin{equation}
{\cal{G}}_{1}(t, r, \theta, u_{r}, u_{\theta}) =
{\cal{A}}(t, r, u_{r}, u_{\theta}) + {\cal{B}}(t, r, u_{r}, u_{\theta}) \, \cos 2\theta +
{\cal{C}}(t, r, u_{r}, u_{\theta}) \, \sin 2\theta \p
\label{ABC}
\end{equation}
This yields the set of equations
\begin{eqnarray}
&& 
\dfrac{\partial {\cal{A}}}{\partial t} + 
u_{r} \, \dfrac{\partial {\cal{A}}}{\partial r} +
\left( \omega^{2} \, r + 2 \, u_{\theta} \, \omega +
\dfrac{u_{\theta}^{2}}{r} 
- \dfrac{\mu}{r^{2}} \right) \dfrac{\partial {\cal{A}}}{\partial u_{r}} 
- \left( 2 \, u_{r} \, \omega + \dfrac{u_{r} \, u_{\theta}}{r} \right) 
\dfrac{\partial {\cal{A}}}{\partial u_{\theta}} 
\nonumber
\\[15pt]
&& \mbox{} \quad
= 
- \, \dfrac{1}{2} \, r \, \dfrac{\partial {\cal{G}}_{0}}{\partial u_{r}} \sv
\label{equA}
\\[15pt]
&&
\dfrac{\partial {\cal{B}}}{\partial t} + 
u_{r} \, \dfrac{\partial {\cal{B}}}{\partial r} +
\left( \omega^{2} \, r + 2 \, u_{\theta} \, \omega +
\dfrac{u_{\theta}^{2}}{r} 
- \dfrac{\mu}{r^{2}} \right) \dfrac{\partial {\cal{B}}}{\partial u_{r}} 
- \left( 2 \, u_{r} \, \omega +
\dfrac{u_{r} \, u_{\theta}}{r} \right) \dfrac{\partial {\cal{B}}}{\partial u_{\theta}} 
\nonumber
\\[15pt]
&& \mbox{} \quad
= 
- \, 2 \, \dfrac{u_{\theta}}{r} \, {\cal{C}}  -
\dfrac{3}{2} \, r \, \dfrac{\partial {\cal{G}}_{0}}{\partial u_{r}} \sv
\label{equB}
\\[15pt]
&&
\dfrac{\partial {\cal{C}}}{\partial t} + 
u_{r} \, \dfrac{\partial {\cal{C}}}{\partial r} +
\left( \omega^{2} \, r + 2 \, u_{\theta} \, \omega +
\dfrac{u_{\theta}^{2}}{r} 
 - \dfrac{\mu}{r^{2}} \right) \dfrac{\partial {\cal{C}}}{\partial u_{r}} 
- \left( 2 \, u_{r} \, \omega +
\dfrac{u_{r} \, u_{\theta}}{r} \right) \dfrac{\partial {\cal{C}}}{\partial u_{\theta}} 
\nonumber
\\[15pt]
&& \mbox{} \quad
= 
2 \, \dfrac{u_{\theta}}{r} \, {\cal{B}}  +
\dfrac{3}{2} \, r \, \dfrac{\partial {\cal{G}}_{0}}{\partial u_{\theta}} \p
\label{equC}
\end{eqnarray}
Eqs.~\eq{equA}-\eq{equC}, where ${\cal{A}}$, ${\cal{B}}$ and ${\cal{C}}$ are the new unknowns,
do not contain the variable $\theta$; this is the main advantage for solving
the approximate model given by Eqs.~\eq{equA}-\eq{equC} instead of Eq.~\eq{eqGsc}.
\subsubsection{The equation \eq{equA}}
We first consider only this equation, because ${\cal{A}}$ does not appear in
Eqs.~\eq{equB}-\eq{equC}, and we look for exact analytical solutions.
To this scope, we assume that
\begin{equation}
{\cal{G}}_{0} =
\mathbb{G}_{0} \left( \varphi_{1}, \varphi_{2} \right) =
\mathbb{G}_{0} \left(  r \, u_{\theta} + \omega \, r^{2} ,
\dfrac{1}{2} \left( u_{r} \right)^{2} + \dfrac{1}{2} 
\left( u_{\theta} + \omega \, r \right)^{2} - \dfrac{\mu}{r} \right) .
\label{G0A}
\end{equation}
The right hand side of this equation is
$$
- \, \dfrac{1}{2} \, r \, \dfrac{\partial {\cal{G}}_{0}}{\partial u_{r}}
=
- \, \dfrac{1}{2} \, r \, \dfrac{\partial \mathbb{G}_{0}}{\partial u_{r}}
=
- \, \dfrac{1}{2} \, r \,  \dfrac{\partial \mathbb{G}_{0}}{\partial \varphi_{2}} \, u_{r}  
$$
If we define 
\begin{equation}
\mathbb{D}_{A}\left( \varphi_{1}, \varphi_{2} \right) = 
\dfrac{\partial \mathbb{G}_{0}}{\partial \varphi_{2}} \left( \varphi_{1}, \varphi_{2} \right)
\label{DA}
\end{equation}
then, by using again the method of the characteristic curves (see, details in Appendix B),
we prove that the class of functions
\begin{equation}
{\cal{A}} = 
- \, \dfrac{1}{4} \, r^{2} \, \mathbb{D}_{A} \left( \varphi_{1}, \varphi_{2} \right) 
+ \mathbb{A}_{*} \left( \varphi_{1}, \varphi_{2} \right)
\end{equation}
satisfy Eq.~\eq{equA}, provided $\mathbb{A}_{*}$ is differentiable.
%
%
%
%
%
%
%
%
\subsubsection{The system \eq{equB}-\eq{equC}}
We recall the equations
\begin{eqnarray}
&& 
\dfrac{\partial {\cal{B}}}{\partial t} + 
u_{r} \, \dfrac{\partial {\cal{B}}}{\partial r} +
\left( \omega^{2} \, r + 2 \, u_{\theta} \, \omega +
\dfrac{u_{\theta}^{2}}{r} 
- \dfrac{\mu}{r^{2}} \right) \dfrac{\partial {\cal{B}}}{\partial u_{r}} 
- \left( 2 \, u_{r} \, \omega +
\dfrac{u_{r} \, u_{\theta}}{r} \right) \dfrac{\partial {\cal{B}}}{\partial u_{\theta}} 
\nonumber
\\[15pt]
&& \mbox{} \quad
= 
- \, 2 \, \dfrac{u_{\theta}}{r} \, {\cal{C}}  -
\dfrac{3}{2} \, r \, \dfrac{\partial {\cal{G}}_{0}}{\partial u_{r}} \sv
\\[15pt]
&&
\dfrac{\partial {\cal{C}}}{\partial t} + 
u_{r} \, \dfrac{\partial {\cal{C}}}{\partial r} +
\left( \omega^{2} \, r + 2 \, u_{\theta} \, \omega +
\dfrac{u_{\theta}^{2}}{r} 
 - \dfrac{\mu}{r^{2}} \right) \dfrac{\partial {\cal{C}}}{\partial u_{r}} 
- \left( 2 \, u_{r} \, \omega +
\dfrac{u_{r} \, u_{\theta}}{r} \right) \dfrac{\partial {\cal{C}}}{\partial u_{\theta}} 
\nonumber
\\[15pt]
&& \mbox{} \quad
= 
2 \, \dfrac{u_{\theta}}{r} \, {\cal{B}}  +
\dfrac{3}{2} \, r \, \dfrac{\partial {\cal{G}}_{0}}{\partial u_{\theta}} \p
\end{eqnarray}
Now, we look for solutions of this kind
\begin{eqnarray}
&&
{\cal{B}} =
\mathbb{B}(r, \varphi_{1}, \varphi_{2}) =
\mathbb{B}\left( r, r \, u_{\theta} + \omega \, r^{2} ,
\dfrac{1}{2} \left( u_{r} \right)^{2} + \dfrac{1}{2} 
\left( u_{\theta} + \omega \, r \right)^{2} - \dfrac{\mu}{r} \right) ,
\\[12pt]
&&
{\cal{C}} =
u_{r} \, \mathbb{C}(r, \varphi_{1}, \varphi_{2}) =
u_{r} \, \mathbb{C}\left( r, r \, u_{\theta} + \omega \, r^{2} ,
\dfrac{1}{2} \left( u_{r} \right)^{2} + \dfrac{1}{2} 
\left( u_{\theta} + \omega \, r \right)^{2} - \dfrac{\mu}{r} \right) .
\end{eqnarray}
It is easy to verify that the equations become
\begin{eqnarray}
&&
u_{r} \, \dfrac{\partial \mathbb{B}}{\partial r} +
2 \, \dfrac{u_{\theta}}{r} \, u_{r} \, \mathbb{C} = \mbox{} -
\dfrac{3}{2} \, r \, \dfrac{\partial {\cal{G}}_{0}}{\partial u_{r}} \sv
\label{BG}
\\[12pt]
&&
(u_{r})^{2} \, \dfrac{\partial \mathbb{C}}{\partial r} +
\left( \omega^{2} \, r + 2 \, u_{\theta} \, \omega +
\dfrac{u_{\theta}^{2}}{r} - \dfrac{\mu}{r^{2}} \right) \mathbb{C}
- 2 \, \dfrac{u_{\theta}}{r} \, \mathbb{B} =
\dfrac{3}{2} \, r \, \dfrac{\partial {\cal{G}}_{0}}{\partial u_{\theta}} \p
\label{CG}
\end{eqnarray}
Now
$$
\dfrac{u_{\theta}}{r} = \dfrac{\varphi_{1}}{r^{2}} - \omega 
\sv \qquad
(u_{r})^{2} = 2 \, \varphi_{2} - \dfrac{\varphi_{1}^{2}}{r^{2}} + 2 \, \dfrac{\mu}{r} 
\sv \qquad
\omega^{2} \, r + 2 \, u_{\theta} \, \omega + \dfrac{u_{\theta}^{2}}{r} = 
\dfrac{\varphi_{1}^{2}}{r^{3}} \sv
$$
and, since we have assumed that
$\dm {\cal{G}}_{0} = \mathbb{G}_{0} \left( \varphi_{1}, \varphi_{2} \right)$,
we also have
$$
r \, \dfrac{\partial {\cal{G}}_{0}}{\partial u_{\theta}} =
r^{2} \, \dfrac{\partial \mathbb{G}_{0}}{\partial \varphi_{1}} + \varphi_{1} \,
\dfrac{\partial \mathbb{G}_{0}}{\partial \varphi_{2}} \p
$$
Therefore, if we define
$$
\mathbb{D}_{B}(\varphi_{1}, \varphi_{2}) = 
\dfrac{\partial \mathbb{G}_{0}}{\partial \varphi_{1}}(\varphi_{1}, \varphi_{2}) \sv
$$
then the system \eq{BG}-\eq{CG} writes
\begin{eqnarray}
&& 
\dfrac{\partial \mathbb{B}}{\partial r} +
2 \left( \dfrac{\varphi_{1}}{r^{2}} - \omega \right) \mathbb{C} =
- \, \dfrac{3}{2} \, r \, \mathbb{D}_{A} \left( \varphi_{1}, \varphi_{2} \right) ,
\label{Br}
\\[12pt]
&&
\left( 2 \, \varphi_{2} - \dfrac{\varphi_{1}^{2}}{r^{2}} + \dfrac{2 \, \mu}{r} \right)
\dfrac{\partial \mathbb{C}}{\partial r} +
\left( \dfrac{\varphi_{1}^{2}}{r^{3}} - \dfrac{\mu}{r^{2}} \right) \mathbb{C}
- 2 \left( \dfrac{\varphi_{1}}{r^{2}} - \omega \right) \mathbb{B} =
\nonumber
\\[12pt]
&& \mbox{} \qquad
\dfrac{3}{2} \left( r^{2} \, \mathbb{D}_{B} \left( \varphi_{1}, \varphi_{2} \right) + 
\varphi_{1} \, \mathbb{D}_{A} \left( \varphi_{1}, \varphi_{2} \right) \right) .
\label{Cr}
\end{eqnarray}
We note that the two equations are simple ordinary differential equations, where $r$ is the only
variable, because $\varphi_{1}$, $\varphi_{2}$ will play only the role of parameters.
Eqs.~\eq{Br}-\eq{Cr} can be solved numerically with suitable initial conditions,
using standard routines.
%
%
%
%
%
%
%
\section{Appendix A}
We consider the set of ordinary differential equations

\begin{equation}
\left\{
\begin{array}{l}
\dfrac{dr}{dt} = u_{r}
\\[12pt]
\dfrac{d \theta}{dt} = \dfrac{u_{\theta}}{r}
\\[12pt]
\dfrac{d u_{r}}{dt} = \omega^{2} \, r + 2 \, u_{\theta} \, \omega + \dfrac{u_{\theta}^{2}}{r} 
 - \dfrac{\mu}{r^{2}} 
\\[12pt]
\dfrac{d u_{\theta}}{dt} = - \, 2 \, u_{r} \, \omega - \dfrac{u_{r} \, u_{\theta}}{r}
\end{array}
\right. .
\label{eqchG0bis}
\end{equation}
From the last equation of Eq.~\eq{eqchG0bis}, taking into account the first equation, we derive
$$
\dfrac{d u_{\theta}}{dt} + \, 2 \, \omega \, \dfrac{dr}{dt} + \dfrac{u_{\theta}}{r} \, 
\dfrac{dr}{dt} = 0
\quad \Leftrightarrow \quad
\dfrac{d (r \, u_{\theta})}{dt} + \omega \, \dfrac{d \, r^{2}}{dt}  = 0 \p
$$
Hence, we have
\begin{equation}
r \, u_{\theta} + \omega \, r^{2} = k_{1} \sv
\label{intk1}
\end{equation}
where $k_{1}$ is a constant.
Now, we consider Eq.~\eq{eqchG0bis}$\mbox{}_{3}$.
Using Eq.~\eq{intk1}, we have
$$
\dfrac{d u_{r}}{dt} = \omega^{2} \, r + 2 \left( \dfrac{k_{1}}{r} - \omega \, r \right) \omega + 
\dfrac{1}{r} \left( \dfrac{k_{1}}{r} - \omega \, r \right)^{2} - \dfrac{\mu}{r^{2}}
\quad \Leftrightarrow \quad
\dfrac{d u_{r}}{dt} = \dfrac{k_{1}^{2}}{r^{3}} - \dfrac{\mu}{r^{2}} \p
$$
Taking into account Eq.~\eq{eqchG0bis}$\mbox{}_{1}$, we obtain
$$
\dfrac{d^{2} r}{dt^{2}} = \dfrac{k_{1}^{2}}{r^{3}} - \dfrac{\mu}{r^{2}}
\quad \Leftrightarrow \quad
\dfrac{d \mbox{ }}{dt} \left[ \dfrac{1}{2} \left( \dfrac{dr}{dt} \right)^{2} \right] =
\dfrac{d \mbox{ }}{dt} \left( - \, \dfrac{1}{2} \, \dfrac{k_{1}^{2}}{r^{2}} + \dfrac{\mu}{r}
\right) ,
$$
that is
$$
\dfrac{1}{2} \left( \dfrac{dr}{dt} \right)^{2} +
 \dfrac{1}{2} \, \dfrac{k_{1}^{2}}{r^{2}} - \dfrac{\mu}{r} = k_{2} \sv
$$
where $k_{2}$ is another constant.
It is useful to eliminate the constant $k_{1}$ by means of Eq.~\eq{intk1} and to use
Eq.~\eq{eqchG0bis}$\mbox{}_{1}$.
We obtain
\begin{equation}
\dfrac{1}{2} \left( u_{r} \right)^{2} + \dfrac{1}{2} 
\left( u_{\theta} + \omega \, r \right)^{2} - \dfrac{\mu}{r} = k_{2} \p
\label{intk2}
\end{equation}
The physical meaning of Eqs.~\eq{intk1}-\eq{intk2} is clear, because the equations of the 
characteristic curves describes the motion of a particle under the influence of the gravitational 
force of a central body in a comoving frame.
To solve the full system \eq{eqchG0bis}, we can use this simple flowchart.
\begin{enumerate}
\item 
We solve the equation 
$ \dm 
\dfrac{1}{2} \left( \dfrac{dr}{dt} \right)^{2} +
 \dfrac{1}{2} \, \dfrac{k_{1}^{2}}{r^{2}} - \dfrac{\mu}{r} = k_{2}
$
and we find $r(t)$.
\item
Since $ \dm u_{r} = \dfrac{dr}{dt}$ we obtain $u_{r}(t)$ by differentiating.
\item
The equation $ \dm u_{\theta} = \dfrac{k_{1}}{r} - \omega \, r$ gives $u_{\theta}(t)$,
immediately.
\item
We solve the differential equation $ \dm \dfrac{d \theta}{dt} = \dfrac{u_{\theta}}{r}$ to have
$\theta(t)$.
\end{enumerate}

\section{Appendix B}
We consider Eq.~\eq{equA}. Taking into account \eq{DA}, it writes 
$$
\dfrac{\partial {\cal{A}}}{\partial t} + 
u_{r} \, \dfrac{\partial {\cal{A}}}{\partial r} +
\left( \omega^{2} \, r + 2 \, u_{\theta} \, \omega +
\dfrac{u_{\theta}^{2}}{r} 
 - \dfrac{\mu}{r^{2}} \right) \dfrac{\partial {\cal{A}}}{\partial u_{r}} 
- \left( 2 \, u_{r} \, \omega + \dfrac{u_{r} \, u_{\theta}}{r} \right) 
\dfrac{\partial {\cal{A}}}{\partial u_{\theta}} 
= 
- \, \dfrac{1}{2} \, r \, u_{r} \, \mathbb{D}_{A} \p
$$
The method of the characteristic curves furnish the set of differential equations
$$
\left\{
\begin{array}{l}
\dfrac{dr}{dt} = u_{r}
\\[12pt]
\dfrac{d u_{r}}{dt} = \omega^{2} \, r + 2 \, u_{\theta} \, \omega + \dfrac{u_{\theta}^{2}}{r} 
 - \dfrac{\mu}{r^{2}}
\\[12pt]
\dfrac{d u_{\theta}}{dt} = - \, 2 \, u_{r} \, \omega - \dfrac{u_{r} \, u_{\theta}}{r}
\\[12pt]
\dfrac{d {\cal{A}}}{dt} = - \, \dfrac{1}{2} \, r \, u_{r} \, \mathbb{D}_{A} \p
\end{array}
\right.
\quad
\Leftrightarrow
\quad
\left\{
\begin{array}{l}
\dfrac{dr}{dt} = u_{r}
\\[12pt]
\dfrac{1}{2} \left( u_{r} \right)^{2} + \dfrac{1}{2} 
\left( u_{\theta} + \omega \, r \right)^{2} - \dfrac{\mu}{r} = k_{2}
\\[12pt]
r \, u_{\theta} + \omega \, r^{2} = k_{1}
\\[12pt]
\dfrac{d {\cal{A}}}{dt} = - \, \dfrac{1}{2} \, r \, u_{r} \, \mathbb{D}_{A}(k_{1}, k_{2}) \p
\end{array}
\right. \p
$$
The first equation gives
$$
\dfrac{dr}{dt} = u_{r} \Rightarrow
r \, \dfrac{dr}{dt} = r \, u_{r} \Rightarrow
r \, u_{r} = \dfrac{1}{2} \,  \dfrac{d r^{2}}{dt} \p
$$
Hence, the last equation becomes
$$
\dfrac{d {\cal{A}}}{dt} = - \, \dfrac{1}{4} \,  \dfrac{d r^{2}}{dt} \, \mathbb{D}_{A}(k_{1}, k_{2})
\qquad \Leftrightarrow \qquad
{\cal{A}} + \, \dfrac{1}{4} \, r^{2} \, \mathbb{D}_{A}(k_{1}, k_{2}) = k_{3} \p
$$
Therefore, it is simple matter to verify that
$$
{\cal{A}}(t, r, u_{r}, u_{\theta}) =
- \, \dfrac{1}{4} \, r^{2} \, \mathbb{D}_{A} \left( \varphi_{1}, \varphi_{2} \right) 
+ {\cal{A}}_{*}(t, r, u_{r}, u_{\theta}) 
$$
satisfies Eq.~\eq{equA}, provided ${\cal{A}}_{*}$ is a solution to the equation
$$
\dfrac{\partial {\cal{A}}_{*}}{\partial t} + 
u_{r} \, \dfrac{\partial {\cal{A}}_{*}}{\partial r} +
\left( \omega^{2} \, r + 2 \, u_{\theta} \, \omega +
\dfrac{u_{\theta}^{2}}{r} 
 - \dfrac{\mu}{r^{2}} \right) \dfrac{\partial {\cal{A}}_{*}}{\partial u_{r}} 
- \left( 2 \, u_{r} \, \omega + \dfrac{u_{r} \, u_{\theta}}{r} \right) 
\dfrac{\partial {\cal{A}}_{*}}{\partial u_{\theta}} 
= 0 \sv
$$
which is Eq.~\eq{eqG0} again.
%
%
%
%
%
%
%
%

%

\begin{thebibliography}{99}
%
%
\bibitem{Batt}
Batt, J., Faltenbacher, W. and Horst, E.:
Stationary spherically symmetric models in stellar dynamics. 
Arch. Ration. Mech. Anal. {\bf 93}, 159-183 (1986)
%
\bibitem{Binney}
Binney, J. and Tremaine, S:
Galactic Dynamics, Princeton Univ. Press, Princeton, New York (1987)
%
\bibitem{Cheng}
Cheng, Y. and Gamba, I.M.:
Numerical study of one-dimensional Vlasov-Poisson equations
for infinite homogeneous stellar systems.
Commun. Nonlinear Sci. Numer. Simul. {\bf 17}, 2052-2061 (2012)
%
\bibitem{Griv2000}
Griv, E., Gedalin, M., Eichler, D. and Yuan, C.:
A gas-kinetic stability analysis of self-gravitating and collisional
particulate disks with application to Saturn's rings.
Planet. Space Sci. {\bf 48}, 679-698 (2000)
%
\bibitem{Griv2003a}
Griv, E., Gedalin, M. and Yuan, C.:
On the stability of Saturn's rings: a quasi-linear kinetic theory.
Mon. Not. R. Astron. Soc. {\bf 342}, 1102-1116 (2003)
%
\bibitem{Griv2003b}
Griv, E., Gedalin, M. and Yuan, C.:
On the stability of Saturn's rings to gravity disturbances.
A\&A {\bf 400}, 375-383 (2003)
%
\bibitem{Griv2003c}
Griv, E. and Gedalin, M.:
The fine-scale spiral structure of low and moderately high optical depth
regions of Saturn's main rings: A review.
Planet. Space Sci. {\bf 51}, 899-927 (2003)
%
\bibitem{Rein}
Rein, G.:
Collisionless kinetic equations from astrophysics – the Vlasov-Poisson system.
Handbook of differential equations: evolutionary equations. Vol. III, 383-476,
Elsevier/North-Holland, Amsterdam (2007)
%
\bibitem{Severne}
Severne, G. and Haggerty, M.J.:
Kinetic theory for finite inhomogeneous gravitational systems.
Astrophys. Space Sci. {\bf 45}(2), 287-302 (1976) 
%
\bibitem{Yoshikawa}
Yoshikawa, K., Yoshida, N. and Umemura, M.: 
Direct integration of the collisionless Boltzmann equation in six-dimensional phase space:
Self-gravitating systems.
Astrophys. J. {\bf 762}(2), art. no. 116. doi:10.1088/0004-637X/762/2/116 (2013) 
%
\end{thebibliography}
\end{document}